\documentstyle[aps,preprint,eqsecnum]{revtex}
\global\parskip = 2pt plus 0.2pt% paragraph skip

\def\endignore{}
\def\ignore #1\endignore{}% use to "comment out" text

\ifx\epsffile\undefined
\def\insertfig#1{}% null macro
\else
\def\insertfig#1{{\baselineskip=4pt
\centerline{\epsfxsize=\hsize\epsffile{#1}}}}\fi

\def\beq{\begin{equation}}
\def\eeq{\end{equation}}

\def\beqa{\begin{eqnarray}}
\def\eeqa{\end{eqnarray}}

\def\eq#1{(\ref{#1})}

\def\hepph#1{{\tt hep-ph/#1}}

\def\jref#1#2#3#4{{\it #1} {\bf #2}, #3 (#4)}

\def\NPB#1#2#3{\jref{Nucl.\ Phys.}{B#1}{#2}{#3}}

\def\PLB#1#2#3{\jref{Phys.\ Lett.}{#1B}{#2}{#3}}
\def\PR#1#2#3{\jref{Phys.\ Rep.}{#1}{#2}{#3}}
\def\PRD#1#2#3{\jref{Phys.\ Rev.}{D#1}{#2}{#3}}

\def\PRL#1#2#3{\jref{Phys.\ Rev.\ Lett.}{#1}{#2}{#3}}

\def\etc{{\em etc\/}}
\def\ie{{\em i.e\/}.}
\def\eg{{\em e.g\/}.}

\def\to{\mathop{\rightarrow}}

\def\myint{\int\mkern-5mu}
\def\sfrac#1#2{{\textstyle\frac{#1}{#2}}}  % small fraction

\def\Dsl{\hbox{\kern.1em/\kern-.7000em$D$}} % D slash

\def\sup#1{\mathstrut^{#1}}

\def\scr#1{{\cal #1}}
\def\op#1{{\widehat #1}}
\def\mybar#1{\kern 0.8pt\overline{\kern -0.8pt#1\kern -0.8pt}\kern 0.8pt}
\def\sla#1{\raise.15ex\hbox{$/$}\kern-.57em #1}% Feynman slash
\def\Sla#1{\kern.15em\raise.15ex\hbox{$/$}\kern-.72em #1}% big Feynman slash

\def\roughly#1{\mathrel{\raise.3ex\hbox{$#1$\kern-.75em%
    \lower1ex\hbox{$\sim$}}}}
\def\lsim{\roughly<}

\def\tr{\mathop{\rm tr}}

\def\al{\alpha}
\def\be{\beta}

\def\De{\Delta}
\def\ep{\epsilon}
\def\Ga{\Gamma}
\def\de{\delta}

\def\La{\Lambda}
\def\si{\sigma}
\def\Si{\Sigma}

\def\Om{\Omega}

\def\ChPT{\raise.45ex\hbox{$\chi$}PT}

\def\hc{{\rm h.c.}}

\def\MeV{{\rm \ MeV}}
\def\GeV{{\rm \ GeV}}

\def\rket#1{|#1)}
\def\rbra#1{(#1|}
\def\op#1{\{#1\}}

\def\sup#1{^{(#1)}}

\def\aem{\al_{\rm EM}}

\begin{document}
\tighten
\preprint{\vbox{
\hbox{MIT-CTP-????}
\hbox{hep-ph/9510453}}}
\title{Baryon masses at second order\\
in large-$N$ chiral perturbation theory}

\author{Paulo F. Bedaque%
\footnote{E-mail: {\tt bedaque@mitlns.mit.edu}}\\
\medskip
Markus A. Luty%
\footnote{Address starting January 1996:
Department of Physics, University of Maryland, College Park MD 20742.
E-mail: {\tt luty@ctp.mit.edu}}
\medskip}

\address{Center for Theoretical Physics\\
Massachusetts Institute of Technology\\
Cambridge, MA 02139\medskip}

\date{October 1995}

\maketitle

\begin{abstract}
We consider flavor breaking in the the octet and decuplet baryon
masses at second order in large-$N$ chiral perturbation theory, where
$N$ is the number of QCD colors.
We assume that
$1/N \sim 1/N_F \sim m_s / \La \gg m_{u,d}/\La, \aem$, where $N_F$
is the number of light quark flavors, and $m_{u,d,s} / \La$ are the
parameters controlling $SU(N_F)$ flavor breaking in chiral
perturbation theory.
We consistently include non-analytic contributions to the baryon
masses at orders $m_q^{3/2}$, $m_q^2 \ln m_q$, and
$(m_q \ln m_q) / N$.
The $m_q^{3/2}$ corrections are small for the relations that follow
from $SU(N_F)$ symmetry alone, but the corrections to the large-$N$
relations are large and have the wrong sign.
Chiral power-counting and large-$N$ consistency allow a 2-loop
contribution at order $m_q^2 \ln m_q$, and a non-trivial explicit
calculation is required to show that this contribution vanishes.
At second order in the expansion, there are eight relations that
are non-trivial consequences of the $1/N$ expansion, all of which
are well satisfied within the experimental errors.
The average deviation at this order is $7 \MeV$ for the $\De I = 0$
mass differences and $0.35 \MeV$ for the $\De I \ne 0$ mass
differences, consistent with the expectation that the error is of
order $1/N^2 \sim 10\%$.
\end{abstract}
\pacs{?}

%--------------------------------------------------------------------
\section{INTRODUCTION}
%--------------------------------------------------------------------
In this paper, we analyze the octet and decuplet baryon masses to
second order in a simultaneous chiral and $1/N$ expansion, where $N$
is the number of QCD colors.
The $1/N$ expansion for baryons has a rich structure and significant
predictive power: at leading order in the $1/N$ expansion, static
baryon matrix elements satisfy $SU(2 N_F)$ spin-flavor relations
(where $N_F$ is the number of flavors) \cite{qm,gs} and the $1/N$
corrections to these relations are highly constrained
\cite{firstucsd,seconducsd,harvard,lmr}.
In order to compare the predictions of the $1/N$ expansion with
experiment at subleading order, we must consider both $1/N$
corrections and explicit breaking of $SU(N_F)$ flavor symmetry due to
quark masses and electromagnetism.
We will use the formalism of refs.~\cite{lmr,nf} that takes $N_F \sim
N \gg 1$ and makes use of an explicit effective lagrangian that keeps
the $SU(N_F)$ flavor symmetry manifest order-by-order in the $1/N$
expansion.
In order to determine which terms in the double expansion to keep, we
will expand assuming that
\beq
\frac 1N \sim \frac 1{N_F}
\sim \frac{m_s}{\La} \gg \frac{m_{u,d}}{\La} \gg \aem,
\eeq
where $\La$ is the chiral expansion parameter and $\aem$ is
the electromagnetic coupling.%
\footnote{See section IIC for a discussion of the last inequality.}
The baryon magnetic moments were analyzed in the same expansion in
ref.~\cite{nmag} and found to be in excellent agreement with
experiment.
Different $1/N$ expansions for various baryon observables have also
been considered in refs.~\cite{seconducsd,ucsdop,jl,djmax}.
In the concluding section we will briefly compare our results with
those of ref.~\cite{jl}, which also considers baryon masses.

One important feature of the present work is that we consistently
include the chiral loop contributions in our expansion.
The leading non-analytic contributions to the baryon mass differences
are order $m_q^{3/2}$, $(m_q \ln m_q) / N$, and $m_q^2 \ln m_q$.
Power-counting and large-$N$ consistency arguments allow a 2-loop
contribution at order $m_q^2 \ln m_q$, and a non-trivial explicit
calculation is required to see that such a contribution does not
appear.
The order $m_q^{3/2}$ corrections are calculable, and the result is
that they are small for the relations that follow from $SU(N_F)$
symmetry alone, but the corrections to the large-$N$ relations are
large and have the wrong sign.
While this may indicate that the expansion parameters are not
sufficiently small in nature to believe this expansion, we note
that there are higher-order effects that are expected to substantially
reduce these corrections.
Also, the $m_q^{3/2}$ corrections can be cancelled at higher orders,
giving results that agree well with experiment.
The coefficients of the $(m_q \ln m_q) / N$ and $m_q^2 \ln m_q$
corrections are not calculable in terms of measured couplings, and we
have included them as arbitrary parameters.

Our final results include corrections up to order $m_s^2$ and $m_s/N$
for the $\De I = 0$ mass differences; and order
$(m_d - m_u) m_s$, $(m_d - m_u)/N$, and $\aem$ for the $\De I \ne 0$
mass differences.
At this order there are eight relations which are non-trivial
consequences of the $1/N$ expansion.
(That is, they do not follow from flavor symmetry considerations
alone.)
These relations agree well with experiment, and the remaining
deviations are consistent with the expectation that the leading
corrections are order $1/N^2$.

This paper is organized as follows.  In section 2, we review the
expansion used in this paper.  In section 3, we present our results,
and section 4 contains our conclusions.  The details of a 2-loop
computation are contained in an appendix.

%--------------------------------------------------------------------
\section{THE EXPANSION}
%--------------------------------------------------------------------
The $1/N$ expansion for baryons has a good deal of predictive power
even at subleading orders, but some aspects of the expansion are
rather subtle.
In this section, we review the main ingredients of this expansion.

%--------------------------------------------------------------------
\subsection{Baryon quantum numbers}
The $1/N$ expansion makes sense only for baryons with spin $J \sim 1$,
since baryons with spin $J \sim N$ have width of order $N$.
One conceptual subtlety in the $1/N$ expansion for baryons is that
even for fixed $N_F \ge 3$ the number of baryon states with given
spin $J \sim 1$ grows with $N$.
When $N_F = 2$, the baryons have quantum numbers
$I = J = \sfrac 12, \sfrac 32, \ldots$, and it is clear that one
should identify the states of lowest spin and isospin with the
corresponding baryon states at $N = 3$.
When $N_F \ge 3$, the situation is more complicated.
The Young tableaux for the $SU(N_F)$ representation of spin $J$
baryons is shown in fig.~1;
for $N > 3$, this representation contains many states that may be
identified with a given baryon state in the world with $N = 3$.
Choosing any subset of the baryon states that exist for $N > 3$ to
represent the baryons at $N = 3$ breaks the flavor symmetry
explicitly.
The ``extra'' states that appear for $N > 3$ are important for
computing chiral loops, since they can appear as intermediate
states;
the contributions of all baryon states are required in order to
maintain flavor symmetry and large-$N$ consistency.

A simple way to handle this situation was pointed out in
refs.~\cite{harvard,lmr}.  The $1/N$ expansion can be carried out
without selecting any subset of the baryon states for $N > 3$ by
writing the matrix elements in terms of few-body operators in a
spin--flavor Fock space that describes the baryon quantum numbers.  In
this formalism, the coefficient of an $r$-body operator is at most
$1/N^{r - 1}$, and there is a simple classification of the operators
that allows us to determine the $N$- and $N_F$-dependence of the
matrix elements for arbitrary baryon states \cite{lmr}.  For operators
with the flavor numbers of a product of $SU(N_F)$ adjoints, any
operator can be written as a product of 1-body operators of the form
\beq
\label{opcan}
\op{X \Ga} \equiv a^\dagger_{a\al} X^a{}_b \Ga^\al{}_\be a^{b\beta},
\eeq
where $a^\dagger$ ($a$) is a creation (annihilation) operator in the
spin--flavor Fock space, $X$ is a flavor matrix and $\Ga$ is a spin
matrix (either 1 or $\si^j$), with flavor and Lorentz indices
contracted in all possible ways.  We keep a given operator if its
matrix element in {\em any} baryon state with $J \sim 1$ (for $N_F \ge
3$) is as large as the order to which we are working.
It is not hard to see that the largest matrix elements are
\beq
\label{matcount}
\op{X \Ga} \lsim \cases{1 & if $X = 1$, $\Ga = \si^j$, \cr
N & otherwise. \cr}
\eeq
Since the operators have definite $SU(N_F)$ quantum numbers, this
procedure keeps the flavor symmetry manifest for arbitrary $N$.

\ignore
\footnote{It should be emphasized that if an operator is large enough
to be important, we evaluate its matrix elements exactly at
$N_F = N = 3$ without dropping terms that are subleading in the $1/N$
expansion.  Dropping subleading terms in the matrix elements would
require us to identify specific large-$N$ states with the physical
baryons, and would break flavor symmetry.  The only ambiguity inherent
in our prescription corresponds to the possibility of adding operators
that are subleading in $1/N$.}
\endignore

\ignore
Using these ideas, one can write an effective lagrangian for the
interactions of baryons and soft pseudo Nambu--Goldstone bosons in
which the $SU(N_F)_L \times SU(N_F)_R$ chiral symmetry and the
large-$N$ counting of matrix elements is manifest \cite{lmr}.
\endignore

%--------------------------------------------------------------------
\subsection{Large $N_F$}
In ref.~\cite{nf} it was shown that the large-$N$ counting rules for
baryons are unaffected when the number of flavors is taken to be
large, \ie\ $N_F \sim N$.  From the point of view of the quark and
gluon degrees of freedom, the reason for this is that the large-$N$
counting rules arise from the suppression of many-quark interactions,
and this suppression is unaffected by the presence of quark loops.
This is to be contrasted with the large-$N$ predictions for mesons,
which generally rely on the suppression of quark loops and therefore
are valid only in the limit $N_F \ll N$.

We will assume that $N_F \sim N \gg 1$.
This seems sensible, since $N = N_F = 3$ in nature.
In order to carry out this expansion, we must decide how to
extrapolate flavor breaking (the quark masses and charges) to a world
with $N_F > 3$.
Clearly, there are infinitely many choices, and our approach will be
similar in spirit to the way we handle the additional baryon states
that occur: we set up the expansion to be independent of the details
of the extrapolation.

We therefore consider extrapolations with arbitrary numbers of
individual quark flavors $N_{u,d,s}$ with $N_u + N_d + N_s = N_F$.
There are extrapolations where $N_q \sim N_F$ ($q = u,d,s$), but there
are also extrapolations where \eg\ $N_s \sim 1$ and $N_{u,d} \sim N_F$.
We will evaluate the matrix elements for arbitrary $N_q$, and keep any
operator that is as large as the order to which we are working on
{\em any} baryon state for {\em any} extrapolation.
Thus for example, we keep terms of order $1/N_q$ as well as $N_q / N$.

%--------------------------------------------------------------------
\subsection{The double expansion}
The predictions of the large-$N$ limit for baryons can be summarized
by stating that baryon matrix elements obey $SU(2 N_F)$
spin--flavor relations in this limit.
In order to consider corrections to this limit, we must take into
account both the fact that $N$ is finite, and the fact that the
$SU(N_F)$ flavor symmetry is broken by quark masses and
electromagnetism.
We therefore carry out a simultaneous expansion in $1/N$ and flavor
breaking.
In order to do this, we must decide where to truncate the expansion
in the small parameters $1/N \sim 1/N_F$, $m_{u,d,s} / \La$, and
$\aem$.

We will expand the $\De I = 0$ mass differences assuming that
\beq
\ep \equiv \frac 1N \sim \frac 1{N_F} \sim \frac{m_s}{\La}.
\eeq
The predictions we obtain will be viewed as predictions for
$\De S \sim 1$ mass differences, which are $O(\ep)$ at leading
order in this expansion.

For the $\De I \ne 0$ mass differences, we assume that
\beq
\aem \ll \frac{m_d - m_u}{\La}.
\eeq
To see that this is reasonable, we note that (to first order in the
expansion performed below)
\beqa
\frac{\aem}{(m_d - m_u)/\La} &\sim&
\frac{m_{\pi^\pm}^2 - m_{\pi^0}^2} {m_{K^0}^2 - m_{K^\pm}^2} = 30\%,
\\
&\sim& \frac{M_{\Si^+} - 2 M_{\Si^0} + M_{\Si^-}}{M_{\Si^+} -
M_{\Si^-}} = 20\%.
\eeqa
We therefore expand $\De I \ne 0$ mass differences to order
$(m_d - m_u)\ep$ and $\aem$.
Except for the fact that we treat $N$ as a large parameter, this
expansion is identical to the second-order chiral expansion that is
usually adopted in chiral perturbation theory \cite{gl,km}.

%--------------------------------------------------------------------
\subsection{Effective lagrangian}
The effective lagrangian we use is described in ref.~\cite{lmr}, and
we will not review it in detail here.
We keep track of the large-$N$ group theory by writing the baryon
fields $\rket B$ as elements in a spin--flavor Fock space.
The operators that couple to these fields are written as $r$-body
operators in spin--flavor space with large-$N$ (and $N_F$) suppression
factors $1/N^{r + t - 1}$, where $t$ is the number of flavor traces
used to write the operator \cite{lmr,nf}.
The fields for the light pseudoscalar mesons $\Pi$ are collected in
the usual combination $\xi \equiv e^{i\Pi / f}$, where
$f \simeq 114 \MeV$ is the kaon decay constant.
(Note that the $\eta'$ mass is of order $N_F / N$, and is therefore not
included as a light field.)
Explicit flavor breaking appears through the quark masses and
electromagnetic charge matrix
\beq
m_q = \pmatrix{m_u {\bf 1}_{N_u} &&\cr & m_d {\bf 1}_{N_d} &\cr
&& m_s {\bf 1}_{N_s} \cr},
\qquad
Q_q = \pmatrix{\frac 23 {\bf 1}_{N_u} &&\cr
& -\frac 13 {\bf 1}_{N_d} &\cr && -\frac 13 {\bf 1}_{N_s} \cr}.
\eeq

%--------------------------------------------------------------------
\subsection{Power counting for chiral loops}
In this subsection, we discuss the power counting of loop graphs in
the expansion described above.
This power counting is more complicated than for ordinary
chiral perturbation theory, and it is essential in order to ensure
that we have not omitted any important contributions at the order
we are working.
(For example, we will see that there are 2-loop graphs that
potentially contribute to the baryon mass differences at order
$m_s^2 \ln m_s$, and we resort to an explicit calculation to see that
it does not occur.)
We begin by reviewing the power counting for ``pure'' baryon chiral
perturbation theory (including only the lowest-lying baryon octet).
We then discuss the new features that are present in the $1/N$
expansion.

In ``pure'' baryon chiral perturbation theory, a generic term in the
effective lagrangian looks like \cite{jm}
\beq
\scr L \sim \sum \La^2 f^2
\left( \frac\partial\La \right)^d
\left( \frac{m_q\vphantom\partial}\La \right)^s
\left( \frac\Pi f \right)^k
\left( \frac B{f\sqrt{\La}} \right)^n
\eeq
where $B$ is the baryon field, and $\Pi$ is the meson field.
(For brevity, we count $\aem Q_q^2$ as a power of $m_q$ in
this subsection.)
A loop diagram will contribute to the baryon mass
\beqa
\de M &\sim& \left( \frac 1{16\pi^2 f^2} \right)^L
\frac 1{\La^{D - 2V_\Pi - V_B}}
\left( \frac{m_q}\La \right) ^S\, m_\Pi^{2L + D - 2V_\Pi - V_B + 1}
\,F(\De M / m_\Pi)
\\
&\sim& \La \left( \frac{\La^2}{16\pi^2 f^2} \right)^L
\left( \frac{m_q\vphantom{\La^2}}{\La\vphantom{f^2}} \right)^C
F(\De M / m_\Pi)
\eeqa
where
\beq
\label{Cdef}
C \equiv L + \sfrac 12 + (\sfrac 12 D_\Pi + S_\Pi - V_\Pi)
+ \sfrac 12 (D_B + S_B - V_B) + \sfrac 12 S_B.
\eeq
Here, $L$ is the number of loops, $D_\Pi$ ($D_B$) is the total number
of derivatives in meson (baryon) vertices, $S_\Pi$ ($S_B$) is the
total number of powers of $m_q$ summed over the meson (baryon)
vertices, $V_\Pi$ ($V_B$) is the total number of pion (baryon)
vertices, and $F(\De M / m_\Pi)$ is a function of the baryon mass
differences and pion masses that appear in the loop diagram.
Because $\De M / m_\Pi \sim m_q^{1/2}$, $F$ can be expanded in a
power series in $m_q^{1/2}$.
The chiral suppression factor $C$ is written in this form because
$D_B + S_B - V_B$ and $\frac 12 D_\Pi + S_\Pi - V_\Pi$ measure the
total number number of ``extra'' insertions of derivatives and/or
powers of $m_q$ in baryon and meson vertices, respectively.
This is because each baryon vertex contains at least one derivative or
one power of $m_q$, and each each pion vertex contains at least two
derivatives or one power of $m_q$.

As an application of this formula, note that at order $m_q^2 \ln m_q$
we must include 1-loop graphs with a single insertion of a
2-derivative baryon vertex ($L = 1$, $S_B = 0$, and
$D_B + S_B - V_B = 1$).
Since the 2-derivative meson vertices are not measured, we cannot
compute the $m_q^2 \ln m_q$ logarithmic corrections to the baryon
masses in baryon chiral perturbation theory \cite{ll}.

We now turn to loop graphs in large-$N$ baryon chiral perturbation
theory.
The chiral suppressions are as given above, and so we will concentrate
on the $N$-dependence of diagrams.
An arbitrary diagram na\"\i vely gives a contribution to the baryon mass
\beq
\label{naive}
\de M \buildrel ? \over \lsim \La
\left( \frac{\La^2}{16\pi^2 f^2} \right)^L
\left( \frac{m_q\vphantom{\La^2}}{\La\vphantom{f^2}} \right)^C
N^{V_B} F(\De M / m_\Pi).
\eeq
Because a meson vertex can change the baryon flavor quantum
numbers only by order 1 (rather than $N$), $\De M$ can have
contributions that are at most $1/N$ or $m_q$.
In our expansion $F(\De M / m_\Pi)$ can therefore be expanded in a
power series in $1/N$ and $m_q^{1/2}$.

Since $f \sim N^{1/2}$, eq.~\eq{naive} apparently violates
large-$N$ consistency if $V_B - L > 1$.
However, there are cancellations among graphs that make the
contributions to the physical mass differences at most order $N$.
Actually, it has never been proven that the required cancellations
occur to all orders, but many explicit calculations have been done
that confirm this assertion \cite{firstucsd,seconducsd}.%
\footnote{In fact, the first few terms in the $1/N$ expansion for
a given matrix element can be derived by demanding that these
cancellations occur \cite{gs,firstucsd}.}
The 2-loop calculation done in the appendix of this paper provides
an additional highly nontrivial check.%
\footnote{2-loop corrections to the axial currents are considered
in ref.~\cite{greg}.}

In our expansion, the argument of $F$ is
\beq
\frac{\De M}{m_\Pi} \sim \frac 1N \,
\left( \frac{\La}{m_q} \right)^{1/2},
\eeq
and we can expand $F$ in a power series in $\De M / m_\Pi$ for loops
of $K$'s and $\eta$'s.
(Loops with $\pi$ intermediate states are negligible at the order we
are working.)
The terms with extra powers of $1/N$ arising from the expansion of
$F$ require fewer cancellations for consistency, and we obtain
\beq
\label{cancel}
\de M \lsim \La
\left( \frac{N \La^2}{16\pi^2 f^2} \right)^L
\left( \frac{m_q\vphantom{\La^2}}{\La\vphantom{f^2}} \right)^C
\left[ N^{\mathop{\rm min}\{V_B - L, 1\}} +
N^{\mathop{\rm min}\{V_B - L - 1, 1\}}
\left( \frac{\La}{m_q} \right)^{1/2} + \cdots \right].
\eeq
The leading contribution to the terms in the square brackets
proportional to $m_q^{n/2}$ arise from diagrams with $n$ insertions
of the $1/N$-suppressed mass term $\De_0$.
The largest term in the brackets in eq.~\eq{cancel} is
\beq
N \left( \frac{\La}{m_q} \right)^{(V_B - L - 1)/2},
\eeq
so that
\beq
\de M_B \le N \La \left( \frac{m_q}{\La}
\right)^{C - (V_B - L - 1)/2}.
\eeq
Using $V_B - L \le L$, we can write
\beq
\label{lastcount}
C - \sfrac{1}{2} (V_B - L - 1) \ge \sfrac{1}{2} (L + 1)
+ (\sfrac{1}{2} D_\Pi + S_\Pi - V_\Pi)
+ \sfrac{1}{2} (D_B + S_B - V_B) + \sfrac 12 S_B.
\eeq

Using these results, we can enumerate all of the graphs that we need
to consider to expand the baryon masses to second order in the
expansion described above.
{}From eq.~\eq{lastcount} we see that we need to consider at most 2-loop
graphs.
Starting from the minimum number of insertions of $\De_0$, it is easy
to see that we need to consider the 1-loop graphs of fig.~2, the
1-loop graphs fig.~3 with zero or one insertion of $\De_0$, and the
2-loop graphs of fig.~4 with one insertion of $\De_0$.
In the appendix it is shown that the leading contribution of the
2-loop graphs cancel, so there is no 2-loop contribution at this
order.

%--------------------------------------------------------------------
\section{EXPANSION OF BARYON MASSES}
%--------------------------------------------------------------------
Using the formalism discussed above, we now turn to the expansion
of the baryon masses.

%--------------------------------------------------------------------
\subsection{Leading order}
We first expand the baryon mass differences to order $\ep$ for the
$\De I = 0$ mass differences and to order $\aem$
and $(m_d - m_u)$ for the $\De I \ne 0$ mass differences.
At this order, there are no chiral loop corrections, and the
expansion is determined by the following terms in the effective
lagrangian:
\beq
\de\scr L = -\rbra B \left[ \De_0 + \De_1 + \cdots \right] \rket B,
\eeq
where
\beqa
\label{leadingone}
\De_{0} &=& \frac{\mu}{N} \op{\si^j} \op{\si^j},
\\
\label{leadingtwo}
\De_{1} &=& a_{1} \op{m}
+ \frac{\aem\La}{4\pi} \left[
b_{1} \op{Q} \op{Q}
+ b_{2} \op{Q\si^j} \op{Q\si^j} \right].
\eeqa
Here,
\beq
m \equiv \frac 12 \left( \xi^\dagger m_q \xi + \hc \right), \quad
Q \equiv \frac 12 \left( \xi^\dagger Q_q \xi^\dagger + \hc \right).
\eeq
Note that the coefficients of the operators $\op{Q}\op{Q}$ and
$\op{Q\si^j}\op{Q\si^j}$ appear to violate the large-$N$ counting
rule discussed above.
The reason is that at the quark level these operators arise from
electromagnetic diagrams such as the one in fig.~5, which are
not suppressed by a factor of $1/N$ from gluon vertices.
However, in order to obtain a good large-$N$ limit, we must demand
that $\aem \lsim 1/N$ so that the electromagnetic Coulomb
energy does not overwhelm the strong binding energy for large $N$.
This also explains why the operator $\op{Q^2}$ is not considered
at this order, since the coefficient of this operator is the same
order as the electromagnetic operators considered above.%
\footnote{In fact, this is a good example of how the $1/N$ expansion
can differ qualitatively from the nonrelativistic quark model.
In the quark model, the operators $\op{Q}\op{Q}$ and $\op{Q^2}$ are
expected to be of the same order, while the spin-symmetry violating
operator $\op{Q\si^j} \op{Q\si^j}$ would be suppressed by $1/m_Q^2$,
where $m_Q$ is the constituent quark mass.
In the constituent quark model, we obtain the additional relation
$\Si^{*-} - \Si^{*+} = \Si^- - \Si^+$ with accuracy $40 \pm 15\%$.
Unfortunately, this is neither sufficiently successful nor
sufficiently unsuccessful to decide whether the quark model works
better than the $1/N$ expansion discussed here.}
At this order we obtain the $\De I = 0$ mass relations
\beqa
\label{GMO}
(\Xi - \Si) - (\Si - N) + \sfrac 32 (\Si - \La) &=& 0,
\qquad (8\%)
\\
\label{done}
(\Xi^* - \Si^*) - (\Si^* - \De) &=& 0,
\qquad (3\%)
\\
\label{dtwo}
(\Om - \Xi^*) - (\Xi^* - \Si^*) &=& 0,
\qquad (7\%)
\\
\label{bstrange}
(\Si - N) - (\La - N) &=& 0,
\qquad (40\%)
\\
\label{dstrange}
(\Si^* - \De) - (\La - N) &=& 0,
\qquad (15\%)
\eeqa
and the $\De I \ne 0$ mass relations
\pagebreak
\beqa
\label{CG}
(\Xi^- - \Xi^0) - (\Si^- - \Si^+) + (n - p) &=& 0,
\qquad (7 \pm 11\%)
\\
\label{deltai}
\De^{++} - 3 \De^+ + 3 \De^0 - \De^- &=& 0,
\\
\label{firsti}
(\Xi^{*-} - \Xi^{*0}) - (\Si^{*-} - \Si^{*+}) + (n - p) &=& 0,
\qquad (2 \pm 22\%)
\\
(\Si^{*+} - 2 \Si^{*0} + \Si^{*-})
- (\Si^+ - 2 \Si^0 + \Si^-) &=& 0,
\qquad (40 \pm 100\%)
\\
\label{lastsurvivor}
(\De^0 - \De^+) - (n - p) &=& 0,
\\
(\Si^{*-} - \Si^{*+}) - (\Si^+ - 2 \Si^0 + \Si^-) - 2 (n - p) &=& 0,
\qquad (3 \pm 7\%)
\\
\label{lasti}
(\De^0 - \De^{++}) + (\Si^{*-} - \Si^{*+}) - 4(n - p) &=& 0,
\qquad (40 \pm 20\%)
%
%\Si^0\La &=& \frac 1{2\sqrt 3} \left[
%(\Si^{*+} - \Si^{*-}) - (\Si^+ - \Si^-) \right].
%
\eeqa
In addition, we can extract the quark mass ratio
\beq
\label{extractrone}
R \equiv \frac{m_s - (m_u + m_d)/2}{m_d - m_u} =
6 \frac{\La - N}{(\Si^+ - \Si^-) - 4(\Si^{*+} - \Si^{*-})}
= 110 \pm 30.
\eeq
Here, the name of a baryon denotes its mass.
%, and $\Si^0\La$ is the
%isospin-violating $\Si^0$--$\La$ mixing mass.
Among the $\De I = 0$ mass relations, eq.~\eq{GMO} (the
Gell-Mann--Okubo relation) and eqs.~\eq{done} and \eq{dtwo} (the
decuplet equal spacing rules) are valid at order $m_s$ independently
of the $1/N$ expansion, while eqs.~\eq{bstrange} and \eq{dstrange}
are consequences of the $1/N$ expansion.
(In this limit, the mass of a baryon with strangeness $-S$ is
proportional to $S$.)
Among the $\De I \ne 0$ mass relations, eq.~\eq{CG} (the
Coleman--Glashow relation) is valid up to corrections of order
$(m_d - m_u) m_s$ independently of the $1/N$ expansion \cite{ll}.
Also, the only corrections to eq.~\eq{deltai} are second
order in isospin breaking independently of the $1/N$ expansion.
The remaining $\De I \ne 0$ relations eqs.~\eq{firsti}--\eq{lasti} as
well as eq.~\eq{extractrone} are consequences of the $1/N$ expansion.

The relations are written as linear combinations of mass differences
that go to zero as $m_s \to 0$ (for the $\De I = 0$ relations) or
$(m_d - m_u) \to 0$, $\aem \to 0$ (for the $\De I \ne 0$ relations).
The accuracy quoted for the relations (where data is available) is
defined as the deviation from zero divided by the average of the
absolute value of the mass differences that appear in the equation.
Defined in this way, all of these relations are na\"\i vely expected to
have corrections of order $\ep \sim 30\%$, except for eq.~\eq{deltai},
which is essentially exact in our expansion.
Note that for the $\De I \ne 0$ mass differences the relations that
hold as a consequence of flavor symmetry alone work better than those
that depend on the $1/N$ expansion, but there is apparently no such
pattern for the $\De I \ne 0$ differences.

The value for $R$ obtained is far from the value $R \simeq 25$ obtained
from an analysis of the light pseudoscalar masses \cite{gl}.
The quoted error only takes into account the experimental uncertainty
of the masses and does not include the theoretical uncertainty from
higher-order corrections.
In view of the large errors in the relations above, we do not take
this value very seriously.

A more objective measure of how well these relations work is obtained
by fitting the mass differences to the parameters given above.
The average deviation in a the best fit is $30 \MeV$ for the
$\De I \ne 0$ mass differences and $0.35 \MeV$ for the $\De I \ne 0$
differences.
(If we omit the model-dependent value for $\De^0 - \De^{++}$, we
get an average deviation of $0.20 \MeV$ for the $\De I \ne 0$
mass differences.)
The fit also gives $R \simeq 90$.

%--------------------------------------------------------------------
\subsection{Chiral loops}
The largest corrections to the leading order results in the expansion
we are performing come from the loop diagrams in fig.~3 and are of
order $N m_s^{3/2}$ for the $\De I = 0$ mass differences and order
$N m_s^{1/2} (m_d - m_u)$ for the $\De I \ne 0$ mass differences.
These diagrams can be evaluated from
\beq
\de M_B = \rbra{B} \op{T_A \si^j} \op{T_A \si^k} \rket{B} \,
\frac{ig^2}{6f^2} \myint \frac{d^4 p}{(2\pi)^4}\,
\frac{p^j p^k}{(p^2 - m_A^2) \, p_0},
\eeq
where $A = K, \eta$, \etc.\ labels the pseudoscalar mass eigenstates
and $T_A$ are the corresponding generators normalized so that
$\tr(T_A T_B) = \de_{AB}$.
The effects of $\pi^0$--$\eta$ mixing are incorporated by using the
generator
\beq
T_\eta = T_8 - \frac{(N_u N_d N_s N_F)^{1/2}}{N_u + N_d}\,
\frac{1}{2R} T_3 + O(1/R^2)
\eeq
for the $\eta$ mass eigenstate, where $R$ is the quark mass ratio
defined in eq.~\eq{extractrone}.
The result can be written
\beq
\de M_B = \rbra{B} J_A \scr O_A \rket{B},
\qquad J_A = \frac{g^2}{16\pi f^2} m_A^3,
\eeq
where
\beqa
\label{firstop}
\scr O_K &=& -(N + N_F - 2 N_s) \op{S}
+ \frac 13 \op{S\si^j} \op{\si^j}
+ \op{S} \op{S}
- \frac 13 \op{S\si^j} \op{S\si^j},
\\
\scr O_\eta &=& \frac{2}{3(N_F - N_s)} \op{S\si^j} \op{\si^j}
- \frac{N_F}{3 N_s (N_F - N_s)} \op{S\si^j} \op{S\si^j}
\nonumber
\\
&& \qquad\qquad +\, \frac{1}{6 R} \left[
\frac{N_s}{N_u + N_d} \op{\tau_3 \si^j} \op{\si^j}
+ \op{\tau_3 \si^j} \op{S \si^j} \right],
\\
\label{lastop}
\scr O_{K^\pm} - \scr O_{K^0} &=& -\frac{N_s}{2} \op{\tau_3}
- \frac{1}{2} \op{\tau_3} \op{S}
+ \frac{1}{6} \op{\tau_3 \si^j} \op{S \si^j}.
\eeqa
Here,
\beq
S \equiv \pmatrix{{\bf 0}_{N_u} &&\cr & {\bf 0}_{N_d} &\cr
&& {\bf 1}_{N_s} \cr}, \qquad
\tau_3 \equiv \frac{2}{N_u + N_d}
\pmatrix{N_d {\bf 1}_{N_u} &&\cr & -N_u {\bf 1}_{N_d} &\cr
&& {\bf 0}_{Ns} \cr}.
\eeq
We do not need expressions for the pion loops, since they are
suppressed by $\sim (m_{u,d} / m_s)^{3/2}$ for $\De I = 0$
quantities, and by $\sim \aem\La / m_{u,d}$ for $\De I \ne 0$
quantities (since the contribution of pion loops is proportional
to $m_{\pi^+} - m_{\pi^0}$, which is purely electromagnetic at first
order in chiral perturbation theory).

Note that eqs.~\eq{firstop} through \eq{lastop} are valid for
arbitrary $N_{u,d,s}$ and $N_F$, and that they have a sensible limit
as $N, N_F \to \infty$ independently of the extrapolation of the
quantities $N_{u,d,s}$.
According to the rules of our expansion, we must keep the full
dependence on $N_{u,d,s}$ and $N_F$ since there are limits where each
of the above terms is important.
The physical results are obtained by simply setting $N_F = 3$,
$N_{u,d,s} = 1$.%
\footnote{Note that for $N_F > 3$ there are ``extra'' pseudoscalar
mesons transforming in the fundamental representation of $SU(N_q)$
that also contribute in the chiral loops.
The masses of these states are determined in terms of the ``physical''
states by $SU(N_F)$ flavor symmetry, and so these contributions are
calculable.
However, these contributions go to zero in the physical limit
$N_F = 3$, and so there is no reason to compute them explicitly.}

Substituting these expressions into the lowest order relations, we
obtain the modified $\De I = 0$ relations
\beqa
(\Xi - \Si) - (\Si - N) + \sfrac 32 (\Si - \La) &=&
\sfrac{4}{3} J_K - J_\eta
\nonumber\\
&=& -2 \pm 1 \MeV,
\\
(\Xi^* - \Si^*) - (\Si^* - \De) &=&
\sfrac{4}{3} J_K - J_\eta
\nonumber\\
&=& -2 \pm 1 \MeV,
\\
(\Om - \Xi^*) - (\Xi^* - \Si^*) &=&
\sfrac{4}{3} J_K - J_\eta
\nonumber\\
&=& -2 \pm 1 \MeV,
\\
(\Si - N) - (\La - N) &=&
-\sfrac{4}{3} (J_K + J_\eta)
\nonumber\\
&=& -400 \pm 160 \MeV
\\
(\Si^* - \De) - (\La - N) &=&
\sfrac{2}{3} (J_K + J_\eta)
\nonumber\\
&=& 200 \pm 80 \MeV
\eeqa
and the modified $\De I \ne 0$ relations
\beqa
(\Si^{*-} - \Si^{*+}) - (\Si^+ - 2 \Si^0 + \Si^-) - 2 (n - p) &=&
-\frac{4}{3} (J_{K^0} - J_{K^+}) + \frac{2}{3R} J_\eta
\nonumber\\
&&\kern-10em
= (-4.0 \pm 1.6) + (4.8 \pm 1.9) \left( \frac{24}{R} \right)
\quad (0.1 \pm 0.6 \MeV)
\\
(\De^0 - \De^{++}) + (\Si^{*-} - \Si^{*+}) - 4(n - p) &=&
-\frac{4}{3} (J_{K^0} - J_{K^+}) + \frac{2}{3R} J_\eta
\nonumber\\
&&\kern-10em
= (-4.0 \pm 1.6) + (4.8 \pm 1.9) \left( \frac{24}{R} \right)
\quad (1.6 \pm 0.85 \MeV)
\eeqa
where the errors in the theoretical prediction are obtained by
assigning a $20\%$ uncertainty to the coupling $g = 0.8$ extracted
from a lowest-order fit to the $\De S = 1$ semileptonic baryon decays,
and the experimental values are shown in parenthesis.
(The remaining $\De I \ne 0$ relations do not receive corrections at
this order.)
In the $\De I = 0$ relations, the $O(m_s^{3/2})$ corrections to the
relations that follow from $SU(N_F)$ flavor symmetry alone are
small due to an ``accidental'' cancellation.
(This was noted in ref.~\cite{jmass}.)
However, even taking into account the theoretical errors, the
corrections to the relations that follow from the large-$N$ expansion
are too large and have the wrong sign.
The same conclusion does not appear to hold for the corrections to the
$\De I \ne 0$ relations, although the situation is obscured by the
large uncertainties involved.
This situation is analogous to what happens for the magnetic moments,
where the $O(m_s^{1/2})$ corrections to the large-$N$ relations are
too large, although of the right sign \cite{nmag}.

One can view this situation in several ways.
The most conservative view is that these results show that the
expansion we are performing does not work well.
We will instead adopt the attitude that the apparently large
corrections at this order of the expansion may be misleading, and go
on to higher orders of the expansion.
At higher orders, these large loop contributions can be cancelled by
counterterms, and we will see that the predictions work very well.

To see why our viewpoint may be reasonable, we note that there are
several calculable higher-order effects that we have omitted in our
calculation, all of which substantially suppress the correction.
The first of these are higher-order corrections to the meson coupling
$g$.
These are known to be large and apparently reduce $g$ \cite{jm,lwhite}.
Second, we have checked that using the exact kinematics for the
particles in the loops decreases the loop corrections by approximately
a factor of $\frac 12$ for the large-$N$ relations.
Third, we expect that the meson--baryon coupling decreases at
large momentum transfer, reducing the effects of $K$ and $\eta$ loops.
All of these effects are suppressed by additional powers of $m_s$, and
are therefore higher order in the expansion we are performing.
Although none of these corrections changes the sign of the loop
corrections, it is possible that these suppressions are large enough
that the expansion we are performing makes sense at higher orders.
We certainly cannot prove conclusively that this point of view is
correct, but we can obtain evidence for it by going to higher orders.

There are also nonanalytic corrections of order $m_q^2 \ln m_q$.
These are formally larger than the analytic corrections due to the
counterterms by $\ln \La^2 / M_K^2 = 1.4$ for $\La = 1 \GeV$.
While these effects are definitely enhanced for $m_s$ sufficiently
small, we believe that for the physical value of $m_s$ this enhancement
is not numerically large enough to give reliable predictions without
including the analytic counterterms.

%--------------------------------------------------------------------
\subsection{Second order}
In order to expand the $\De I = 0$ mass differences to order
$\ep^2$ and the $\De I \ne 0$ mass differences to order
$\aem$ and $(m_d - m_u) \ep$, we must include the additional
terms
\beq
\De_2 =
\frac{a_2}N \op{m\si^j} \op{\si^j}
+ \frac{a_3}{\La} \op{m^2}
+ \frac{a_4}{N\La} \op{m} \op{m}
+ \frac{a_5}{N\La} \op{m\si^j} \op{m\si^j}.
\eeq

In addition to the analytic contributions, there are nonanalytic
contributions from loop graphs of order
$m_q^2 \ln m_q$ and $(m_q \ln m_q) / N$.
There are $m_q^2 \ln m_q$ contributions from graphs such as fig.~2
with the baryon--meson vertices coming from higher-order operators
such as
\beq
\frac{1}{\La} \op{A^j A^j}, \quad
\frac{1}{N\La} \op{A^j} \op{A^j}, \quad
\frac{1}{N\La} \op{A^j \si^j} \op{A^k \si^k}.
\eeq
There are also $m_q^2 \ln m_q$ contributions from graphs such
as fig.~3 with one of the baryon--meson vertices coming from
the higher-order operators
\beq
\frac{1}{\La} \op{(\nabla_0 A^j) \si^j}, \quad
\frac{1}{\La} \op{(\nabla_j A^0) \si^j}.
\eeq
The coupling constants associated with these operators are not
well-measured, and we will treat them as free parameters.
These graphs give contributions to the baryons masses of the form
\beq
\de M_B =
\rbra{B} C_j \scr{O}_{jA} K_A \rket{B},
\qquad K_A \equiv \frac{1}{16 \pi^2 f^2 \La} m_A^4
\ln\frac{\La^2}{m_A^2},
\eeq
where the $C_j$ are unknown constants and the independent operators
that contribute at this order are
\beq
\scr{O}_{1A} = \op{T_A} \op{T_A}, \qquad
\scr{O}_{2A} = \op{T_A \si^j} \op{T_A \si^j}.
\eeq
There are also $(m_q \ln m_q) / N$ corrections arising from graphs
such as fig.~3 with a single insertion of the flavor-independent
baryon mass operator $\De_0 = \mu \op{\si^j} \op{\si^j} / N$.
These contributions are proportional to
\beqa
\frac{g^2}{16\pi^2 f^2} [ \op{T_A \si^j}, &&\!\!\!\!\!\!\!\!
[\op{T_A \si^j}, \De_0]]\,
m_A^2 \ln\frac{\La^2}{m_A^2}
\nonumber
\\
&\propto& \frac{g^2}{16\pi^2 f^2}\,
\frac{\mu}{N} \left[
\op{T_A \si^j} \op{T_A \si^j} - \op{T_A T_A \si^j} \op{\si^j}
\right] m_A^2 \ln\frac{\La^2}{m_A^2}
\\
&\sim& \frac{1}{N^2} \left[ N^2 + N N_F \right] m_q \ln m_q.
\eeqa
The normalization of these contributions is calculable, but will
not be needed.

When we expand the operators above in terms of the
$SU(N_F)$-violating spurions $S$ and $\tau_3$, we see that the
operators that appear in $\scr{O}_1$ and $\scr{O}_2$ are linear
combinations of the operators that appear in the tree terms.
However, the loop contributions are important because they
eliminate a prediction for the quark mass ratio $R$ that would
otherwise exist at this order.

Eliminating the unknown constants leads to the relations
\beqa
\label{improveq}
(\Om - \Xi^*) - 2 (\Xi^* - \Si^*) + (\Si^* - \De) &=& 0,
\qquad (1\%) \\
\label{trueone}
(\Xi^* - \Si^*) - (\Si^* - \De)
- (\Xi - \Si) + (\Si - N) - \sfrac{3}{2} (\Si - \La) &=& 0,
\qquad (6\%)
\\
\label{truetwo}
(\Xi^* - \Si^*) - (\Xi - \Si) &=& 0,
\qquad (17\%)
\\
\label{CGagain}
(\Xi^- - \Xi^0) - (\Si^- - \Si^+) + (n - p) &=& 0,
\qquad (7 \pm 11\%)
\\
\De^{++} - 3 \De^+ + 3 \De^0 - \De^- &=& 0,
\\
(\Xi^{*-} - \Xi^{*0}) - (\Si^{*-} - \Si^{*+}) + (n - p) &=& 0,
\qquad (2 \pm 22\%)
\\
(\Si^{*+} - 2 \Si^{*0} + \Si^{*-})
- (\Si^+ - 2 \Si^0 + \Si^-) &=& 0,
\qquad (40 \pm 100\%)
\\
\label{lastold}
(\De^0 - \De^+) - (n - p) &=& 0,
\\
(\De^0 - \De^{++}) + (\Si^+ - 2\Si^0 + \Si^-) - 2 (n - p) &=& 0,
\qquad (70 \pm 30\%)
\eeqa
Of the $\De I = 0$ relations, eq.~\eq{improveq} is an improved
version of the equal spacing rule that holds to order $m_s^2$
independently of the $1/N$ expansion \cite{jmass} while
eqs.~\eq{trueone} and \eq{truetwo} are non-trivial predictions of
the $1/N$ expansion.
These relations were derived in this expansion in ref.~\cite{nf};
the same relations are derived in a different expansion in
ref.~\cite{seconducsd}.
Of the $\De I \ne 0$ relations, eqs.~\eq{CGagain} through \eq{lastold}
are identical to the lowest-order relations eqs.~\eq{CG} through
\eq{lastsurvivor}.
It is worth noting that the Coleman--Glashow relation eq.~\eq{CGagain}
receives calculable {\em analytic} corrections of order
$(m_d - m_u) m_s$ in chiral perturbation theory \cite{ll};
the results above show that these corrections are suppressed by
$1/N$.

A fit to the measured mass differences gives an average deviation
of $7 \MeV$ for the $\De I = 0$ mass differences (compared to $29 \MeV$
at lowest order).
For the $\De I \ne 0$ mass differences, the average deviation is
$0.35 \MeV$, or $0.20 \MeV$ if we exclude the $\De^0$--$\De^{++}$
mass difference.
Since the average of the $\De I = 0$ mass differences is approximately
$150 \MeV$ and the average of the $\De I \ne 0$ mass differences is
approximately $3.5 \MeV$, this is consistent with the expected
accuracy of $1/N^2 \sim 10\%$.
Note that the result for the $\De I \ne 0$ mass difference is the same
as at first order, but this simply reflects the fact that the
first-order relations work better than expected.

\subsection{Higher orders}
At next order in large-$N$ chiral perturbation theory, we must include
the operators
\beqa
\De_3 &=& \frac{a_6}{N^2} \op{m} \op{\si^j} \op{\si^j}
+ \frac{a_7}{N^2\La} \op{m} \op{m \si^j} \op{\si^j},
\nonumber\\
&& \quad  +\, \frac{a_8}{N^2\La^2} \op{m} \op{m} \op{m}
+ \frac{a_9}{N^2 \La^2} \op{m} \op{m\si^j} \op{m\si^j}
\eeqa
which give contributions to the mass differences of order
$m_q / N^2$, $m_q^2 / N$, and $m_q^3$.
Including these terms in addition to the second-order effects
discussed above, there is one surviving relation:
\beq
(\Si^{*+} - 2 \Si^{*0} + \Si^{*-})
= (\Si^+ - 2 \Si^0 + \Si^-).
\qquad (40 \pm 100\%)
\eeq
This relation gets corrections from the term
\beq
\frac{\aem \La}{4\pi}\, \frac{1}{N\La} \op{Q}\op{Q\si^j}\op{m\si^j}
\eeq
at order $\aem m_s$, and also possibly from loop effects at order
$m_q^{5/2}$, $m_q^3 \ln m_q$, \etc.
(Checking this requires us to consider 3-loop graphs.)
In view of the large experimental uncertainty in this relation, we
will not pursue the expansion beyond this point.

%--------------------------------------------------------------------
\section{CONCLUSIONS}
%--------------------------------------------------------------------
We have analyzed the baryon mass differences in large-$N$ baryon
chiral perturbation theory with particular emphasis on the chiral
loop corrections.
One result of our work is that the nonanalytic $m_s^{3/2}$ corrections
to the $\De I = 0$ large-$N$ mass relations appear to be too large and
have the wrong sign to explain the corrections to the lowest-order
relations.
Nonetheless, accurate results are obtained at higher order in this
expansion.
At second order, there are eight non-trivial predictions of the
$1/N$ expansion (which do not follow from flavor symmetry alone).
These relations work to better than $10\%$ accuracy, consistent
with the assumption that the errors are
$\sim 1/N^2 \sim m_s^2 / \La^2$.

Ref.~\cite{jl} also analyzes the baryon masses in a combined
expansion in $1/N$ and flavor breaking.
The main differences from the present work is that ref.~\cite{jl}
performs a different expansion in which only baryons with strangeness
of order 1 in the large-$N$ limit are considered, and nonanalytic
chiral loop corrections are not included.
Loop corrections in the expansion of ref.~\cite{jl} are considered
in ref.~\cite{jleff}.

%--------------------------------------------------------------------
\section*{APPENDIX: BARYON MASSES TO TWO LOOPS}
%--------------------------------------------------------------------
In this Appendix, we compute the 2-loop contributions to the baryon
masses shown in fig.~4.
(The other two-loop graphs can be seen to be negligible using the
power-counting arguments of section IIIB.)
We will find that these contributions are negligible, but we do not
know any simpler way to see this than by computing them explicitly.

%--------------------------------------------------------------------
\subsection*{General Formula}
We begin by deriving the general formula for the 2-loop contribution
to the masses in the presence of mixing.
The baryon self-energy can be viewed as an operator
$\Ga(E)$ in the spin-flavor Fock space, where $E$ is the energy of the
baryon.  If we denote the physical baryon fields by $\rket{n}$, the
mass eigenvalues are determined by
\beq
\label{shell}
\Ga(E_n) \rket{n} = 0.
\eeq
This is a nonlinear eigenvalue equation to be solved simultaneously
for $E_n$ and $\rket{n}$ order by order in the loop expansion.
The first step is to expand all quantities in the number of loops:
\beqa
\Ga(E_n) &=& E_n - \De - L_1(E_n) - L_2(E_n) + \cdots,
\\
E_n &=& E_n\sup 0 + E_n\sup 1 + E_n\sup 2 + \cdots,
\\
\rket{n} &=& \rket{n\sup 0} + \rket{n\sup 1} + \rket{n\sup 2} + \cdots.
\eeqa
The energy dependence of the loop perturbations must also be expanded:
\beq
L_j(E_n) = L_j(E_n\sup 0) + L'_j(E_n\sup 0) E_n\sup 1 + \cdots.
\eeq
Writing out eq.~\eq{shell} and equating terms at the same order in the
loop expansion gives at tree level
\beq
\De \rket{n\sup 0} = E_n\sup 0 \rket{n\sup 0},
\eeq
at one loop%
\footnote{We are assuming that all degeneracies are lifted by the
tree-level terms, so that we do not need degenerate perturbation
theory.}
\beqa
E_n\sup 1 &=& \rbra{n\sup 0} L_1(E_n\sup 0) \rket{n\sup 0},
\\
\rket{n\sup 1} &=& \sum_{m \ne n} \rket{m\sup 0}
\frac{\rbra{m\sup 0} L_1(E_n\sup 0) \rket{n\sup 0}}
{E_n\sup 0 - E_m\sup 0},
\eeqa
and at two loops
\beqa
\label{twoloop}
E_n\sup 2 &=& \rbra{n\sup 0} L_2(E_n\sup 0) \rket{n\sup 0}
+ E_n\sup 1 \rbra{n\sup 0} L'_1(E_n\sup 0) \rket{n\sup 0}
\nonumber\\
&&\qquad
+ \sum_{m \ne n} \frac{|\rbra{m\sup 0} L_1(E_n\sup 0) \rket{n\sup 0}|^2}
{E_n\sup 0 - E_m\sup 0}.
\eeqa

The ``energy denominator'' terms in the two-loop formula are
irrelevant in the limit where we can ignore mixing,
\beq
\rbra{m\sup 0} L_1(E_n\sup 0) \rket{n\sup 0} = 0
\quad\hbox{for\ $m \ne n$}.
\eeq
For the problem at hand, mixing always violates isospin, so that the
energy denominator terms in eq.~\eq{twoloop} are second order in
isospin violation.
This is true for arbitrary $N$ and $N_F$, as can be seen by
considering the generalized isospin and strangeness discussed in
section IIB as good quantum numbers.
Since the baryons of given spin $J$ form an irreducible representation
of $SU(N_F)$, and because any state in such a representation is
uniquely specified by these quantum numbers, we see that mixing
always violates generalized isospin.
We can then write
\beq
\label{nomix}
E_n\sup 2 = \rbra{n\sup 0} L_2(E_n\sup 0)
+ \sfrac 12 \left[ L_1(E_n\sup 0), L'_1(E_n\sup 0) \right]_+
\rket{n\sup 0}.
\eeq
This formula gives the precise meaning of the graphs in fig.~4.
The anticommutator term can be thought of as arising from wavefunction
renormalization.

%--------------------------------------------------------------------
\subsection*{2-loop Contribution}
We begin by considering the ``true'' 2-loop graphs;
we will consider the 1-loop graphs with counterterm insertions in the
following subsection.
By the power-counting considerations of subsection IIIB, we need only
consider graphs of the form in fig.~4 with a single insertion of
$\De_0$.%
\footnote{We have also evaluated the 2-loop graphs with no
mass insertion and verified that the $N$ dependence is consistent.}
These integrals are regulated using \eg\ dimensional regularization
so that all the manipulations below are well-defined.
This contribution is a sum of terms with momentum integrals of the
form
\beq
\myint \frac{d^4 p}{(2\pi)^4}\, \frac{p^j p^k}{p^2 - m_A^2}
\myint \frac{d^4 q}{(2\pi)^4}\, \frac{q^\ell q^m}{q^2 - m_A^2}
F(p_0, q_0) \propto \de^{jk} \de^{\ell m}
\eeq
by 3-dimensional rotational invariance.
We then obtain
\beqa
\de M_B = \frac{g^4}{36 f^4}
\myint \frac{d^4 p}{(2\pi)^4}\, && \frac{\vec p\,^2}{p^2 - m_A^2}
\myint \frac{d^4 q}{(2\pi)^4}\, \frac{\vec q\,^2}{q^2 - m_B^2}
\nonumber\\
\Biggl[&&
\frac 1{p_0^3 (p_0 + q_0)} \left( A X B B A
+ A B B X A \right)
\nonumber\\
&& + \frac 1{p_0^2 (p_0 + q_0)^2} A B X B A
\nonumber\\
&& + \frac 1{p_0^2 q_0 (p_0 + q_0)} A X B A B
\nonumber\\
&& + \frac 1{p_0 q_0 (p_0 + q_0)^2} A B X A B
\nonumber\\
&& + \frac 1{p_0 q_0^2 (p_0 + q_0)} A B A X B
\nonumber\\
\label{bigtwo}
&& - \left(\frac 1{2 p_0^2 q_0^2} + \frac 1{p_0^3 q_0} \right)
\left( A X A B B + B B A X A \right) \Biggl],
\eeqa
where we have used the abbreviations
\beq
A \equiv \op{T_A \si^j}, \quad
B \equiv \op{T_B \si^k},
\eeq
and \eg
\beqa
AB X BA &\equiv& AB \De_0 BA
- \sfrac 12 \left( \De_0 ABBA + ABBA \De_0 \right)
\\
&=& \sfrac 12 [AB, \De_0] BA + \sfrac 12 AB [\De_0, BA],
\eeqa
where $\De_0 = \mu \op{\si^j} \op{\si^j} / N$.

Eq.~\eq{bigtwo} is naively a 5-body operator (if we take into
account the commutator structure defined by $X$ insertions), but it
must be a 4-body operator by large-$N$ consistency.
By explicit calculation, we find that the 5-body part of
eq.~\eq{bigtwo} indeed vanishes, and we obtain
\beqa
\de M_B = \frac{g^4}{36 f^4}
\myint \frac{d^4 p}{(2\pi)^4}\, && \frac{\vec p\,^2}{p^2 - m_A^2}
\myint \frac{d^4 q}{(2\pi)^4}\, \frac{\vec q\,^2}{q^2 - m_B^2}
\nonumber\\
\Biggl(&&
\frac 1{2 p_0^2 q_0^2} A B [A, [B, \De]]
\nonumber\\
&&\quad +\, \frac{4 q_0 + 3 p_0}{2 p_0^3 q_0 (p_0 + q_0)}
A B [[A, B], \De]
\nonumber\\
&&\quad +\, \frac{2 p_0^3 + 3 p_0^2 q_0 + 2 p_0 q_0^2 + 2 q_0^3}
{p_0^3 q_0^3 (p_0 + q_0)} A [A, B] [B, \De]
\Biggr)
\nonumber\\
&&\quad +\, \hbox{lower-body\ operators}.
\eeqa
Evaluating the 4-body operators that appear, we find that they
are all $\lsim N^2$, and we have checked that all lower-body
operators that appear are also $\lsim N^2$ even for $N_F \sim N$.
This shows that
\beq
\de M_B \lsim N^0 m_q^2 \ln m_q,
\eeq
which is negligible in our expansion.

Note that there is a 4-body operator that could have contributed,
namely
\beqa
A A [B, [B, \De]] &\sim&
\frac{1}{N} A A \left(
[B, \op{\si^j}] [B, \op{\si^j}]
+ \op{\si^j} [B, [B, \op{\si^j}]]
+ \cdots \right)
\nonumber
\\
&\sim& \frac{1}{N} N^2 [N^2 + N_F N + \cdots].
\eeqa
Therefore, the important point of this calculation is that this
operator does not appear in the evaluation of the 2-loop graphs.

%--------------------------------------------------------------------
\subsection*{One loop counterterm graphs}
As long as we are evaluating 1-loop graphs, we can use dimensional
regularization and minimal subtraction, which allow us to simply
drop the divergent parts.
When we consider 2-loop graphs, we must be more careful about the
counterterm structure, since we must also evaluate the 1-loop
graphs with a single insertion of a 1-loop counterterm.

As already stated above, diagrams with $L$ loops generally scale as
$N^L$ for large $N$.
This means that there are divergences in 1PI graphs that cannot be
cancelled by counterterms with the same $N$-dependence as the
tree-level lagrangian.
Even at 1 loop, there is a divergent wavefunction renormalization
of order $N$, while the tree-level kinetic term is order $1$.
This does not imply that the lagrangian is not closed under
renormalization, since we can rescale the fields so that the
counterterms have the correct form.%
\footnote{Again, this has not been proven to all loops.}
We therefore write the (bare) effective lagrangian as
\beq
\scr L = \rbra B Z^{1/2} \left[ iv\cdot\nabla - \De +
g \op{A\cdot\si}  + \cdots + \scr O_{\rm ct} \right] Z^{1/2} \rket B,
\eeq
where $Z$ and $\scr O_{\rm cr}$ are divergent counterterms to be
chosen order-by-order in the loop expansion to render all graphs
finite.
By allowing $Z$ to have arbitrary $N$ dependence, we can
cancel the divergences with counterterms $\scr O_{\rm ct}$ that have
the same form as the operators appearing in the tree-level lagrangian.
This is the sense in which the bare lagrangian has the same
form as the tree-level lagrangian.

At one loop, we find
\beqa
\label{zone}
Z &=& 1 - \frac{g^2}{6f^2} \op{T_A \si^j} \op{T_A \si^j} + O(\De^2)
\\
\label{octone}
\scr O_{\rm ct} &=& -\frac{g^2}{12 f^2}
[ \op{T_A \si^j}, [\op{T_A \si^j}, \De - g \op{A\cdot\si} ]]
\mathop{\rm div} \myint \frac{d^4 p}{(2\pi)^4}\,
\frac{\vec p\,^2}{p^2 - m_A^2}\, \frac 1{p_0^2}.
\eeqa
Here, ``div'' indicates the divergent part.

The counterterms in $\scr O_{\rm ct}$ are  smaller by one power of
$1/N$ than required by large-$N$ consistency.
This can be used to see that the contribution of the 1-loop counterterm
graphs to the baryon masses are $\sim N^0$ (not $\sim N$).
The reason is simply that the physical mass is independent of the
scale of the fields, and so one can compute the masses with $Z = 1$.
All the other counterterms that appear are now suppressed by an
additional power of $1/N$, which is immediately gives the result
that these contributions are negligible in our expansion.

%--------------------------------------------------------------------
\section*{ACKNOWLEDGEMENTS}
%--------------------------------------------------------------------
We would like to thank G.L. Keaton for useful discussions.
P.F.B. and M.A.L. thank the theory group at Brookhaven National
Laboratory, and M.A.L. thanks the Aspen Center for Physics and the
theory group at Lawrence Berkeley National Laboratory for hospitality
while this work was in progress.
This work is supported in part by DOE contract DE-AC02-76ER03069 and
by NSF grant PHY89-04035.

%--------------------------------------------------------------------
% REFERENCES
%--------------------------------------------------------------------

%--------------------------------------------------------------------
% FIGURE CAPTIONS
%--------------------------------------------------------------------
\vfill\eject
\centerline{\large FIGURE CAPTIONS}
\bigskip

\noindent
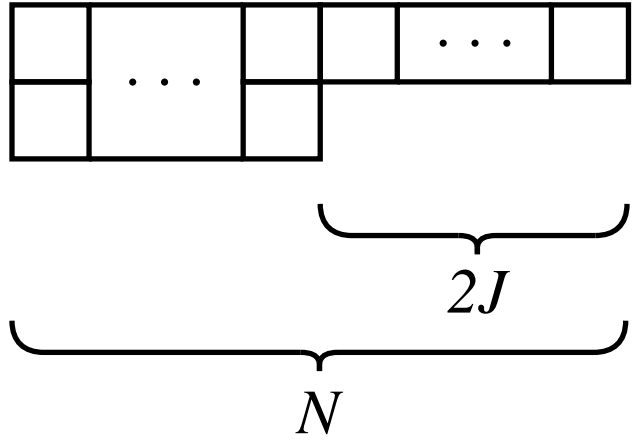
FIG.~1:
The Young tableaux for the $SU(N_F)$ flavor representation
of the spin-$J$ baryon multiplet.
\medskip

\noindent
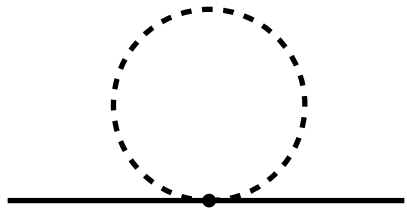
FIG.~2:
A contribution to the baryon mass.  The solid lines are
baryons and the dashed lines are mesons.
\medskip

\noindent
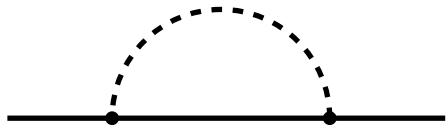
FIG.~3:
A contribution to the baryon mass.
\medskip

\noindent
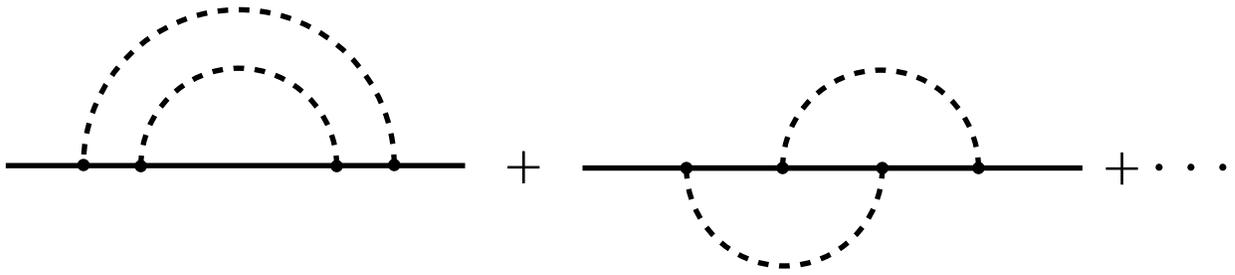
FIG.~4:
Two-loop contributions to the baryon mass.
\medskip

\noindent
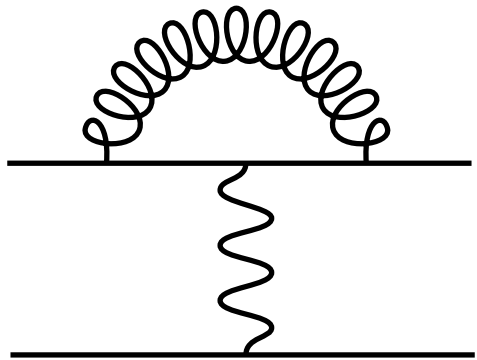
FIG.~5:
Example of a quark graph giving rise to the 2-body operators
$\op{Q}\op{Q}$ and  $\op{Q \si^j} \op{ Q \si^j}$ in the effective
lagrangian.
The wavy line represents a photon and the curly line represents a
gluon.
Note that there is a factor of $N$ from the color sum, so this graph
is order 1 in the large-$N$ limit.
\medskip

\end{document}